\documentclass{PoS}
\title{Comparison of Improved Perturbative Methods}

\ShortTitle{Comparison of Improved Perturbative Methods}

\author{Y. Meurice $^a$, and \speaker{Haiyuan Zou} $^a$\\

$^a$Department of Physics and Astronomy, The University of Iowa\\
Iowa City, Iowa 52242, USA\\
         E-mail: \email{yannick-meurice@uiowa.edu}\\
        E-mail: \email{haiyuan-zou@uiowa.edu}\\}

\abstract{In many cases of interest, the perturbative series based on conventional Feynman diagrams have a zero radius of convergence. Series with a finite radius of convergence can be obtained by either introducing a large field cutoff or by replacing the exponential of the perturbation by a sequence of approximants as recently proposed by Pollet, Prokof'ev, and Svistunov. We compare these two methods for integrals and quantum mechanical problems. The two methods perform well in complementary regime (strong coupling for the large field cutoff and intermediate coupling for the other method). 
We briefly discuss potential applications for lattice gauge theory with compact groups (which have a build-in large field cutoff).}

\FullConference{ The XXIX International Symposium on Lattice Field Theory - Lattice 2011\\
July 10-16, 2011\\
Squaw Valley, Lake Tahoe, California}

\begin{document}

\section{Introduction}
More than 50 years ago, Dyson argued that the radius of convergence of the expansions of physical quantities for QED in powers of $e^2$ is zero because of the vacuum unstability if $e^2$ is changed to $-e^2$ \cite{dyson52}. 
Later, it was established that for many models, the perturbative expansions calculated from Feynman diagrams are divergent \cite{leguillou90}. For instance, for several models with $\lambda \phi^4$ interactions, it is well-known that the perturbative coefficients increase at a factorial rate with the order. The failure of the convergence of the expansion can be understood easily from the simple integral where 
\begin{equation}
\label{eq:instability}
\int_{-\infty}^{+\infty}d\phi e^{-\frac{1}{2}\phi^2-\lambda \phi^4}\neq \sum_{q=0}^{\infty}
\frac{(-\lambda)^q}{q!} \int_{-\infty}^{+\infty}d\phi e^{-\frac{1}{2}\phi^2}\phi^{4q}\ .
\end{equation}
In this case, Dyson's instability is related to large field configurations \cite {convpert,optim03}. 

Several modifications of the large field configurations have been proposed. One possibility consists in truncating the large field contributions \cite{convpert}: in the partition function of  $\lambda \phi^4$ models, the integrals over $(-\infty,\infty)$  are replaced by integrals over $(-\phi_{max},\phi_{max})$. With this modification, Eq (\ref{eq:instability}) becomes an equality, but we are dealing with a different integral than the one we started with.
On the other hand, for lattice gauge models with compact groups, there is a build-in large field cutoff and conventional series are obtained by ``adding the tails 
of integration" \cite{optim05}. 
 
Another modification consists in replacing the original partition function by a convergent sequence of approximations where the exponential of the perturbation 
is replaced by a suitable sum of Bessel functions, as recently proposed by Pollet, Prokof'ev, and Svistunov \cite {pps}. In the following,  this method is abbreviated as the PPS method. As we will see, in this case, the contributions from the large fields have damped oscillation, and the series for the modified integrals become convergent.  

In this proceedings, we compare these two improved perturbative methods mostly for the simple integral $\lambda \phi^4$ theory and 
mention recent progress for the anharmonic oscillator. 

\section{Two Modified Perturbative Methods}
For  models with $\lambda \phi^4$ intereactions, and the perturbative expansion of the generating function  of $n$-point functions can be written as a product of two exponents:
\begin{equation}
\label{eq:partitionf1}
Z[j]\sim e^{-\lambda \sum_{x}(\frac{\partial}{\partial j_{x}})^4} e^{\frac{1}{2}\sum_{y,z}j_{y}G_{y,z}j_{z}} ,
\end{equation}
In the large field cutoff method, the second exponential is modified. In the PPS method, the first exponential is modified. 

\subsection{The Large Field Cutoff Method}
When a large field cutoff is introduced, the standard shifts that allow to solve the free theory with sources need to be modified by writing the truncated integral as the integral over the whole real axis minus the integral over $|\phi|>\phi_{max}$. We then obtain an expension of the form \cite{convpert}:

\begin{equation}
\label{eq:partitionf2}
Z[j]\sim e^{-\lambda \sum_{x}(\frac{\partial}{\partial j_{x}})^4} e^{\frac{1}{2}\sum_{y,z}j_{y}G_{y,z}j_{z}}(1-C \sum_{y}\int_{\phi_{max}}^{+\infty}d\phi_{y}e^{-D(\phi_{y}-\sum_{z}G_{y,z}j_{z})^2}+...)\  .
\end{equation}
If we want to write Feynman rules, the additional terms require a generalization of Wick theorem. Up to now, the modified coefficients have only been calculated numerically. 

\subsection{The PPS Method}
In the PPS method, the vertex part is modified into a new sequence of approximants with a regulator $N$ \cite{pps}. The partition function becomes
\begin{eqnarray}
\label{eq:partitionf3}
\nonumber
Z[j]\rightarrow Z_{N}[j]&=&\int_{-\infty}^{+\infty}d\phi e^{-\frac{1}{2}{\phi}^2}[f({\lambda}^{\frac{1}{4}}\phi,N)]^N\\ 
&\simeq & \sum_{k=0}^{p}{\lambda}^k C_{k} ,
\end{eqnarray}
In which,
$f({\lambda}^{\frac{1}{4}}\phi,N)=J_{0}(a(N){\lambda}^{\frac{1}{4}}\phi)
+2J_{2}(a(N){\lambda}^{\frac{1}{4}}\phi)
+\frac{5}{3} J_{4}(a(N){\lambda}^{\frac{1}{4}}\phi)$\ ,
with $a(N)$ is chosen so that the series of $f({\lambda}^{\frac{1}{4}}\phi,N)$ has the form $1-\frac{\lambda{\phi}^4}{N}+O(\frac{{\lambda}^2{\phi}^8}{N^2})$.

By increasing the regulator $ N$, the modified partition function becomes more like the original one. Note that for the simple integral, $|C_{k}|$ is near-Gaussian and decays rapidly for large $k$. For a fixed $\lambda$, the numerical value of the modified partition function is convergent as the order $p$ increases. Feynman rules can be written easily:
\begin{equation}
\label{eq:partitionf4}
Z[j]\sim e^{-[ \lambda \sum_{x}(\frac{\partial}{\partial j_{x}})^4+b(N)(\frac{{\lambda}^2}{N}\sum_{x}(\frac{\partial}{\partial j_{x}})^8)+\dots ]} e^{\frac{1}{2}\sum_{y,z}j_{y}G_{y,z}j_{z}}\  . 
\end{equation}
The new vertices of order $\lambda^2 $ and higher can be seen as counterterms. For instance, the lowest order counterterm contributes to the vacuum energy at order  ${\lambda}^2$ through the diagram:

\begin{center}
\begin{picture}(14,15)
\put(17,3){\line(1,3){3.5}}
\put(17,3){\line(3,1){10.5}}
\put(17,3){\line(1,-3){3.5}}
\put(17,3){\line(3,-1){10.5}}
\put(17,3){\line(-1,3){3.5}}
\put(17,3){\line(-3,1){10.5}}
\put(17,3){\line(-1,-3){3.5}}
\put(17,3){\line(-3,-1){10.5}}
\put(17,13.5){\oval(7,7)[t]}
\put(17,-7.5){\oval(7,7)[b]}
\put(27.5,3){\oval(7,7)[r]}
\put(6.5,3){\oval(7,7)[l]}
\end{picture}.
\end{center}

\section{Comparison of Modified Partition Functions in 0-D (simple integrals)}

\begin{figure}[h]
\begin{center}
\includegraphics[width=2.2in]{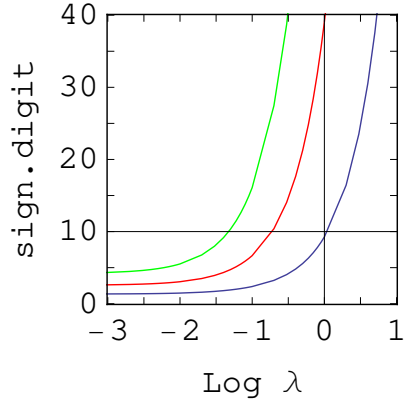}
\end{center}
\caption{\label{fig:errorcut} Number of significant digits for the simple integral obtained from field cutoffs with $\phi_{max} = 2$ (blue), $\phi_{max} = 3$ (red), $\phi_{max} = 4$ (green).}
\end{figure}

\begin{figure}
\begin{center}
 \includegraphics[width=2.2in]{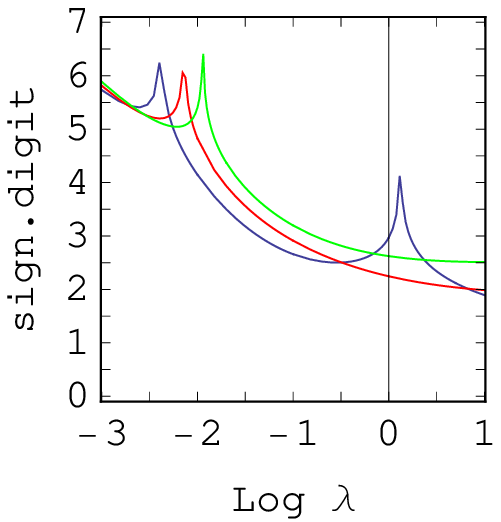}
\includegraphics[width=2.2in]{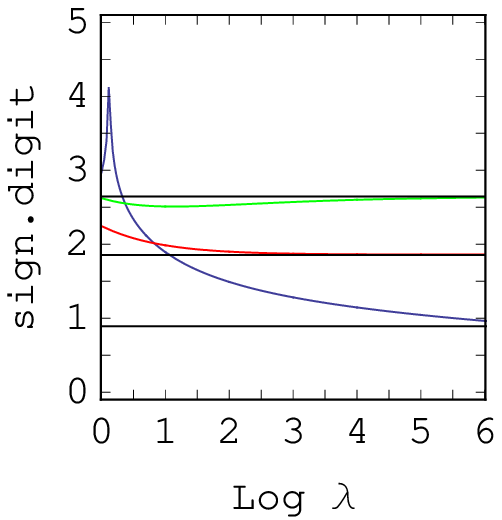}
\end{center}
 \caption{\label{fig:errorpps}Left: Number of significant digits obtained from the PPS modification for the simple integral with regulator $N = 2$ (blue), $N = 3$ (red), $N = 4$ (green). Right: Asymptotic value (black) for the PPS method at large $\lambda$: $N = 2$ (blue), $N = 3$ (red), $N = 4$ (green).}
\end{figure}

In order to assess the quality of various perturbative expansions (or sequences of these),  we ask the two following questions:
\begin{itemize}
\item  
How well does the modified model approximate the original model?
\item
How well does the perturbative series for the modified model at successive orders approximate the modified model?
\end{itemize}
These two questions can be addressed for a broad range of couplings and summarized in what we call ``error graphs'', namely 
the significant digits vs. $log_{10}\lambda$. In all the error graphs, the number of (correct) significant digits is calculated as 
\begin{equation}
-log_{10}|\frac{{\rm modified\  value} -{\rm true\  value}}{{\rm true \  value}}|  \  . 
\end{equation}
In this section, we discuss these two questions for the simple integral (\ref{eq:instability}). We start with the first question. The error graph for the large field cutoff method with different cuts are shown in Fig.\ref{fig:errorcut}. The error introduced by the field cutoff becomes smaller at strong coupling, because in this regime, the large field contribution are strongly suppressed in the original integral. At small coupling, the error graph stabilizes at the error due to removing the tails in the Gaussian limit. 
In comparison, the error graphs for the PPS modified model with different regulator $N$ are shown in Fig. \ref{fig:errorpps} (Left).  We calculated the asymptotic values for the PPS method at small and large $\lambda$. At small $\lambda$, the significant digits are $-log_{10}\frac{21{\lambda}^2}{2N}$.  The asymptotic value for the PPS method at large $\lambda$ can be calculated numerically and are shown in Fig. \ref{fig:errorpps} (right). The methods are complementary in the sense that they have chance to perform well in the opposite limits. 

\begin{figure}
 \begin{center}
 \includegraphics[width=2.1in]{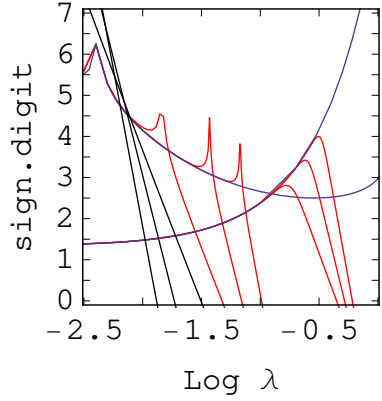}
 \end{center}
 \caption{\label{fig:compare}
Number of significant digits for the simple integral obtained with a field cutoff with $\phi_{max} = 2$ at order 7, 11, 15, as a function of $\lambda$ (red lines with the blue asymptotic line increasing as $\lambda$ increases), from the PPS method with $N=2$ at the same orders (red lines with the blue asymptotic line decreasing as $\lambda$ increases), and from regular perturbative method at the same orders (black).  }
\end{figure}

We can now address the second question regarding the approximation of the modified models by perturbative series.
It seems clear that perturbative expansions of the modified models can at best do as well as the modified models (understood as 
solvable numerically with any desired accuracy).  
 The results for the two modified methods are shown in Fig. \ref{fig:compare}.  One sees that as the order increases, the error of the perturbative expansions gets closer to the 
error due to the modification. In other words, in both cases, the perturbative series converge toward the values taken for the modified models. Fig. \ref{fig:compare} also includes regular perturbation theory and one can see that at sufficiently small coupling it provides the 
best answer. The field cutoff method works well at large coupling and the PPS method at intermediate coupling. 
\section{Logarithms of the Modified Partition Functions in 0-D}

Unlike the perturbative series for the modified partition functions which are convergent for any complex value of $\lambda$,  the expansion of their logarithm have a finite radius of convergence which is the absolute value of the zeros of the modified partitions function closest to the origin in the complex $\lambda$ plane.  
However, accurate values can be obtained by using Pad$\acute{e}$ approximants for both methods. As the order of the Pad$\acute{e}$ approximant increases, the significant digits of the expansion can be improved and we reach the numerical limit
corresponding to the modified models.

\section{Preliminary Results for the Anharmonic Oscillator}

\begin{figure}
 \begin{center}
  \includegraphics[width=2.2in]{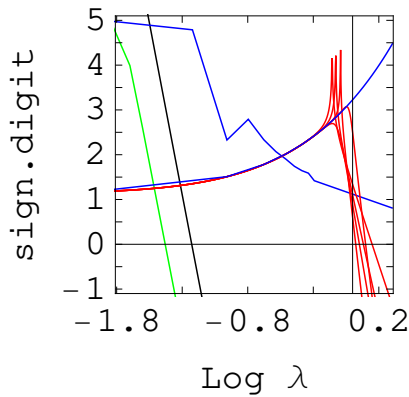}
 \end{center}
  \caption{\label{fig:1dcompare} Error graphs for the anharmonic oscillator. Black: regular perturbation theory at order 15. Blue line (decreasing when $\lambda$ increases): the PPS method at regular $N=4$. Green: the PPS method at order 15. Blue line (increasing when $\lambda$ increases): cutoff method for a cut $=2$. Red: cutoff method at order 9,13,17,21,25.}
\end{figure}

In the last two sections, we found that both the PPS method and the large field cutoff method work in some region of $\lambda$. It is now time to ask if these methods can be used in more complicated models.  In this section, both methods will be extended to the case of the anharmonic oscillator (1D $\lambda \phi^4$). The physical quantity we used to compare with is the ground state energy of the anharmonic oscillator. Following the idea from Ref. \cite{polypot}, we can solve the numerical Schrodinger equation to get the ground state energy for two modified methods and also their perturbative series. However, in the PPS case, the approximation of the potential by polynomials is more subtle because the new modified potential has logarithmic singularities. We are still working on this case, and the preliminary results are shown in Fig. \ref{fig:1dcompare} which shows patterns similar to Fig. \ref{fig:compare}. 

\section{Preliminary Results on the Average Plaquette in U(1) Lattice Gauge Theory}
\begin{figure}
\begin{center}
  \includegraphics[width=2.8in]{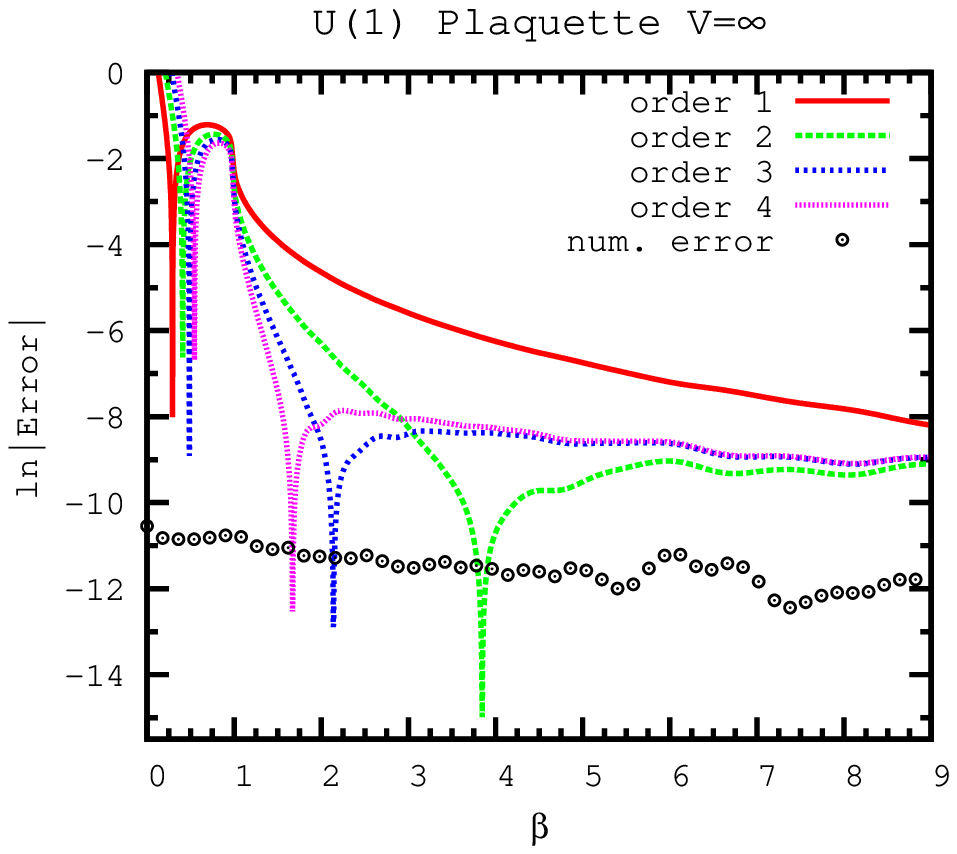}
\includegraphics[width=2.8in]{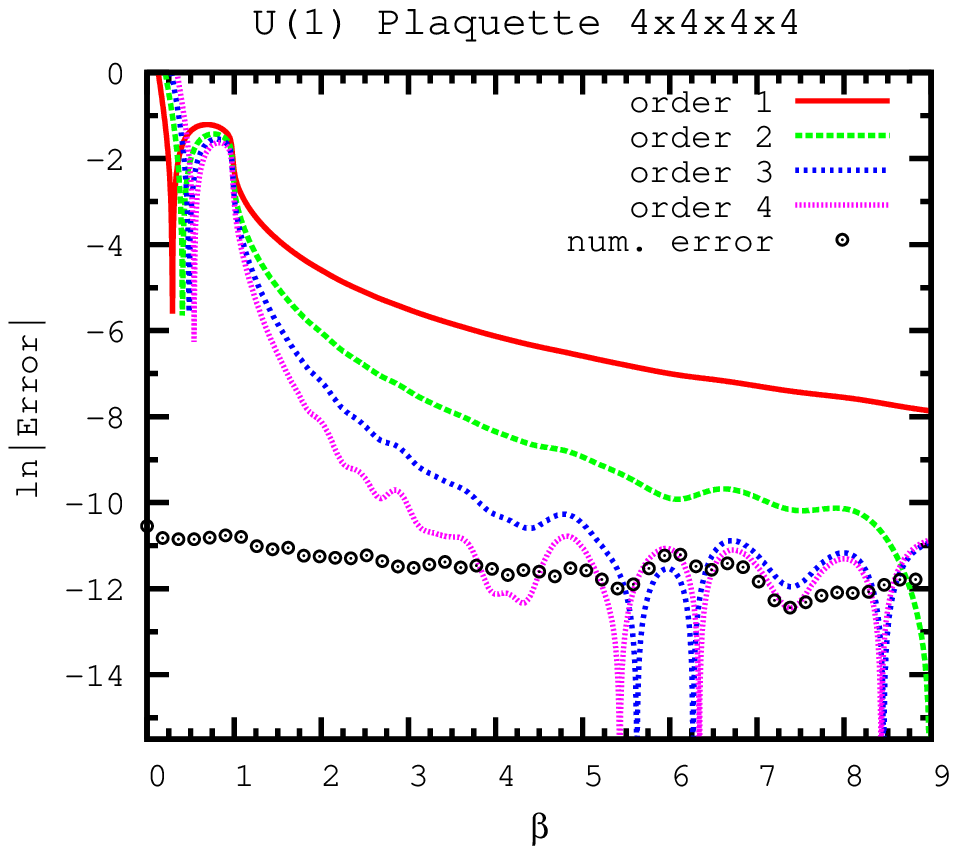}
 \end{center}
  \caption{\label{fig:u1volef} Comparison of the weak coupling expansion at different order with numerical errors without volume correction (left) and with volume correction (right). The numerical error is estimated using a weighted 20-seed data.}
\end{figure}

In order to extend the modified methods to U(1) lattice gauge theory, we need to check the easiest quantity, the average $1\times 1$ plaquette. On the lattice, there are volume effects discussed in Ref. \cite{weakcu1}. We reproduce these volume effects on our error graphs (Fig. \ref{fig:u1volef}), in which the numerical error is estimated through a weighted 20-seed data \cite{ddd}. There is a gap between these two graphs due to the volume corrections. 
The sensitivity to the volume effects indicates that it might be possible to extract a ``perturbative envelope'' as in Ref. \cite{npp} and explain it in terms of large field contributions. 
This is still in progress.

\section{Conclusions}
By introducing a large field cutoff or replacing the exponential of the perturbation by a sequence of approximants, we can get finite radius of convergence for the perturbative series of the logarithms of the simple integrals. The application of these two modified methods in more complicated models, like the anharmonic oscillator and U(1) lattice gauge theory is in progress.

\begin{acknowledgments}
We thank Alexei Bazavov and Daping Du for valuables discussions on U(1) lattice gauge theory and help with Fig. \ref{fig:u1volef}. YM thanks  L. Pollet, N. Prokof'ev, and B. Svistunov for discussions at the Aspen Center for Physics in summer 2010. This research was supported in part by the Department of Energy under Contract No. FG02-91ER40664.

\end{acknowledgments}

\end{document}